# MODULAR COSMIC RAY DETECTOR (MCORD) AND ITS POTENTIAL USE IN VARIOUS PHYSICS EXPERIMENTS, ASTROPHYSICS AND GEOPHYSICS

M. Bielewicz†, M. Kiecana, A. Bancer, J. Grzyb, L.Swiderski, M.Grodzicka-Kobylka,
T. Szczesniak, A. Dziedzic, K. Grodzicki, E. Jaworska, A. Syntfeld-Kazuch,
National Centre for Nuclear Research, Otwock-Swierk, Poland

*Abstract*

As part of the collaboration building a set of detectors for the new collider, our group was tasked with designing and building a large-scale cosmic ray detector, which was to complement the capabilities of the MPD (Dubna) detector set. The detector was planned as a trigger for cosmic ray particles and to be used to calibrate and test other systems. Additional functions were to be the detection of pairs of high-energy muons originating from some particle decay processes generated during collisions and continuous observation of the cosmic muon stream in order to detect multi muons events. From the very beginning, the detector was designed as a scalable and universal device for many applications. The following work will present the basic features and parameters of the Modular COsmic Ray Detector (MCORD) and examples of its possible use in high energy physics, astrophysics and geology. Thanks to its universal nature, MCORD can be potential used as a fast trigger, neutron veto detector, muon detector and as a tool in muon tomography.

## INTRODUCTION

The MCORD detector has been designed as a modular and scalable solution [1]. The basic element is the MCORD section consisting of eight scintillators with built-in optical fiber and double-sided signal readout based on silcon photomultiplayer sensors (SiPM) (16 channels) [2]. The MCORD section, together with its electronics, can work as a stand-alone detection system.

Thanks to the modular design, it is possible to configure detection systems consisting of any number of sections in any geometric arrangement. The size of the scintillators can also be adjusted according to specific requirements. The MCORD section can work independently or in coincident mode.

The detector can be used both in large physics experiments and laboratory measurement stations, and can also be used for educational purposes. Thanks to the hermetic housing of the scintillators, it can also work outdoors in different weather conditions (Fig. 1).

## DETECTOR DESCRIPTION

Full description of the MCORD you can find in his Conceptual Design Report (CDR) document in Ref. [3].

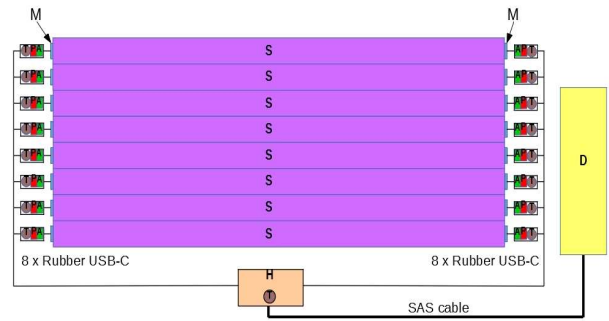

Figure 2: Logical diagram of a single MCORD section.

Each SiPM sensor (reading channel) is equipped with its own temperature sensor, individual amplifier and loops for correcting the supply voltage under the influence of temperature change. Additionally, each detector board (two SiPM sensors) is equipped with its own processor, which allows us to directly monitor and control the operating parameters of each sensor separately. CAN network connectivity with unique ID every chip (SiPM) as CAN address, and unique ID in every hub for cabling and sensor checking and identification.

In Figure 2 we see the functional diagram of a single section of the MCORD detector where: S (purple) – plastic scintillator, M (blue) – SiPM sensor, P (red) – power supply with temperature compensation circuit, T (brown) – temperature sensor, A (green) – amplifier, D (yellow) – MicroTCA system with ADC boards, H (orange) – Passive Signal Hub & Power Splitter, plus set of 8xUSB-C cable (20 Gb/s), 2xSAS-external cable and 1xEthernet cable for one section (8 scintillators).

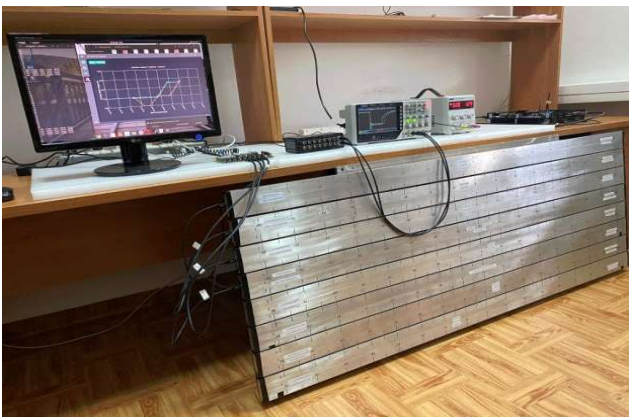

Figure 1: Real photography of single MCORD section.

† marcin.bielewicz@ncbj.gov.pl

Dedicated Analog Front-End module divided into a part installed directly at the scintillators inside their housing and a separate HUB supervising eight detectors (one section) (see Figure 3). A detailed description of the MCORD electronics can be found in Ref. [4].

The scalable digital part for data analysis, based on FPGA circuits and the MTCA platform, enables the construction and creation of software for both simple (1-2 sections) and very complex systems.

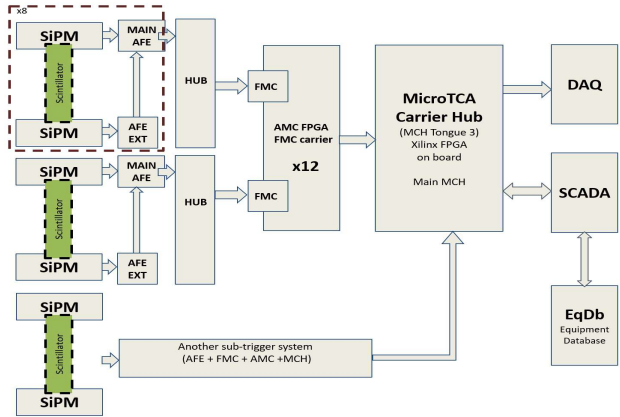

Figure 3: Block diagram of the MCCORD modular electronics system.

### Data Processing

MCORD electronics give us the ability to generate three types of trigger signal depending on needs and available time.

- Latency estimation for L1 trigger (event without parameters): AFE, cabling, ADC + SERDES latency
  Estimated total latency: 1µs
- Latency estimation for L2 trigger (event with parameters): MGT, Algorithm, Formatter and transmitter latency
  Estimated total latency: 3.5 – 7.5µs
- Latency estimation for L3 trigger (between MTCA systems): MGT, Fiber, Algorithm, Formatter and transmitter latency
  Estimated total latency: 10 – 15us

### MCORD Resolution

Position resolution:    X axis – 5-7 cm, Y axis – 7 cm
Time Resolution:        about 700-800 ps
Number of events (particles):  about 100-150 s/m2
Measured efficiency:    99,5 %

### Scintillators, and SiPM/MMPC

- Number of section scintillators:  8 pcs
- Plastic scintillator:             polystyrene (Nuvia)
- WLS fiber:                        2 mm dia. (Kuraray)
- Dimensions of scintillators:      72x22x1620 [mm]
- Dimensions of section:            735x50x1744 [mm]
- Weight of one section:            ~50 kg
- SiPM (MPPC) type:                 3x3 mm (Hamamatsu)

## POTENTIAL MCORD APPLICATIONS IN EXPERIMENTS

Thanks to its modular structure, the MCORD detector can be used in various systems and necessary geometric arrangements. In the simplest case, a single MCORD section can be used as a simple cosmic ray flux monitor. A binary system connected in a coincident system (Figure 4) provides much greater possibilities, from determining the direction to easier particle identification (e.g. detecting muons in the cosmic ray flux). Figure 5 shows several potential situations for using such a system

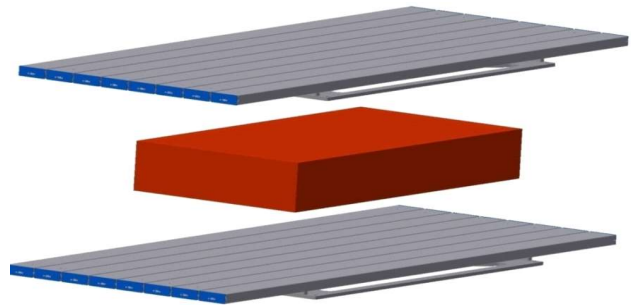

Figure 4: The simplest two sections MCORD coincident detector.

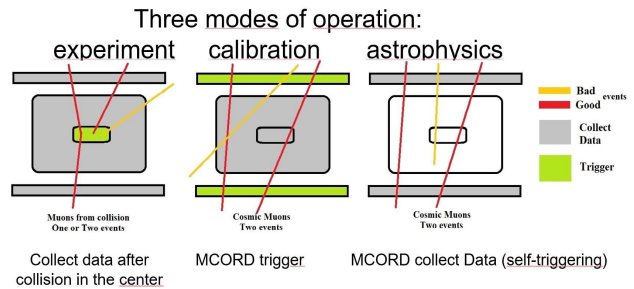

Figure 5: Three example modes of operation of the MCORD detector in coincidence mode, when working with a larger group of other detectors.

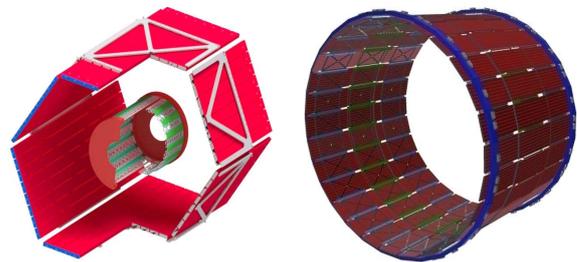

Figure 6: Potential MCORD applications in Experiments. Left: six sections as part of a collision point monitor; Right: 84 sections as a trigger for a large array detectors inside the cylinder.

Figure 6 show two examples of a potential detection system based on MCORD sections. Both cases were designed for real systems in the past. Figure 6 (left) shows an example of using 6 MCORD sections in cooperation with a second smaller detector (which has higher resolution but lower efficiency) to detect and determine the location of the collision point of colliding accelerator beams. Figure 6 (right) shows an example of a possible construction of a large cosmic ray trigger detector based on 84 MCORD sections, which surrounds the entire set of other detectors in the shape of a cylinder. Such a configuration can also be used for astrophysical studies (detection of cosmic rays boundless from any direction).

*Possible applications in physics experiments*

a) Trigger (for testing or calibration):
   - Testing and calibration of different detectors before experimental session (TOF, eCAL, hCAL and TPC).
b) Muon identifier (created after collision with absorber):
   - Pions and Kaons decays,
   - Rare mesons decays (etha, rho),
   - Possible decays of new „dark" particles.
c) Astro physics (muon showers and bundles):
   - unique for horizontal events,
   - working in cooperation with TPC and TOF.
d) Veto detector:
   - veto charged particles for neutral particle detection,
   - gamma detector (cosmic muon veto).

## POTENTIAL USE IN ARCHEOLOGY AND GEOPHYSICS

Detectors of cosmic-ray muons can be used for many very interesting practical applications.

*Muon Tomography for Archeology*

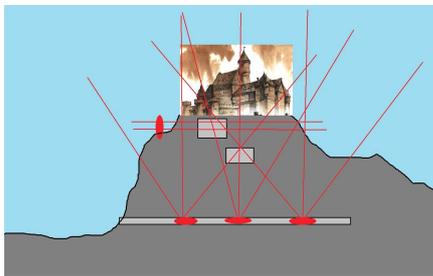

Figure 7: Muon scanning method in archaeology.

The biggest challenge in archaeology is to determine whether there are further empty spaces or spaces of non-standard density under the studied objects. Muon tomography offers the potential for non-invasive and relatively cheap scanning of such areas. Figure 7 shows how to scan areas under the studied object using detectors placed next to the hill or in existing tunnels.

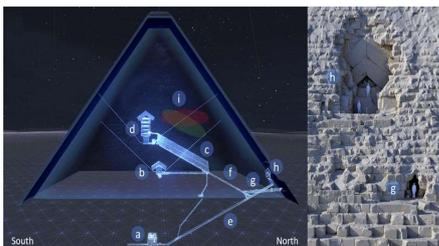

Figure 8: An example of a successful scan of Egyptian pyramids performed by another research group [5].

This method of research has already been used in practice and with great success (see Fig. 8). A group of researchers scanned an Egyptian pyramid using a detector placed inside the pyramid and managed to discover a previously unknown large empty space [5].

*Muon Tomography for Mining*

A Canadian company has developed and promotes the muon scanning method for searching for heavy metal deposits underground [6]. The method uses detectors driven underground at various depths using drilled holes (Fig. 9 left) and, by scanning in many directions, obtains 3D images characterising variation in average material density in a diverse range (Fig. 9 right).

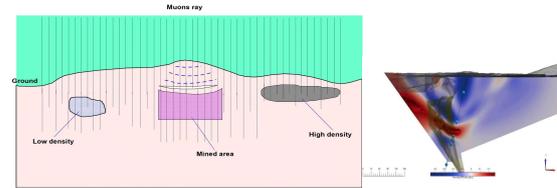

Figure 9: Drilling muon tomography enables the creation of 3D models of underground targets [6].

*Earthquake Precursors*

Large scale correlations between cosmic rays and earthquakes presumably related to earthquakes precursors has been observed by the research group from CREDO collaboration. The found periodicity is rather similar to the sun spots solar cycle. Cosmic ray data correspond to the measurements at the Pierre Auger observatory in Malargüe, Argentina, whereas seismic data is taken from Moscow and Oulu stations located in Russia and Finland, respectively. A 6σ correlation effect has been observed in a period of about 4.5 years. Details can be found in the Ref. [7].

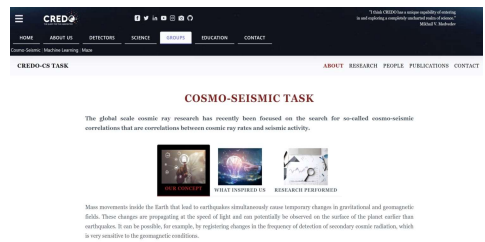

Figure 10: The CREDO page about Cosmo-Seismic task.

Our group, together with the CREDO collaboration, is preparing a program to search for the correlations described above, but on a local scale in Mexico and Chile. We plan to use several MCORD coincidence detectors installed in universities for the creation of a measurement network collecting data simultaneously with seismographic stations [8].

## CONCLUSION

Our group has built a demonstrator of the MCORD cosmic ray detector and performed many measurements and tests of this system. In the above article we have shown the diverse applications of this type of measurement system in many fields of experimental physics and in other fields of science.